\documentclass{aa}  % IRA LaTeX2e version
\usepackage{graphics}
\begin{document}
\thesaurus{04.19.1; 13.18.2}
\title{Multi-Frequency Study of the B3-VLA Sample\footnotemark[1]\\
II. The Database}
\author{ M. Vigotti\inst{1} \and L. Gregorini\inst{1,2} \and  U. Klein\inst{3} 
 \and K.-H. Mack\inst{1,3}}
\institute{Istituto di Radioastronomia del CNR, Via Gobetti 101, I-40129, 
Bologna, Italy
\and Dipartimento di Fisica, Universit\`a di Bologna, Via Irnerio 46, I-40126, 
Bologna, Italy
\and Radioastronomisches Institut der Universit\"at Bonn, Auf dem H\"ugel 71,
D-53121 Bonn, Germany }
\offprints{M. Vigotti (vigotti@ira.bo.cnr.it)}
%\mail{vigotti@ira.bo.cnr.it}
%
%
\date{August 2, 1999}
\maketitle
%
%\markboth{Vigotti et al.}{ }
%\renewcommand{\thefootnote}{\fnsymbol{footnote}}
\footnotetext[1]{Table 2 is also available in electronic
from at the CDS via anonymous ftp to cdsarc.u-strasbg.fr (130.79.128.5) or via 
http://cdsweb.u-strasbg.fr/Abstract.html}
\begin{abstract}
We present total flux densities of 1049 radio sources in the frequency range
from 151~MHz to 10.6~GHz. These sources belong to the B3-VLA sample, which is 
complete down to 100~mJy at 408~MHz. The data constitute a homogeneous spectral 
database for a large sample of radio sources, 50 times fainter than the 3C 
catalogue, and will be used to perform a spectral ageing analysis, which is one 
of the critical points in understanding the physics and evolution of 
extragalactic 
radio sources.
 
\end{abstract}

\section{Introduction} 

Homogeneous databases over a wide frequency range for a large sample of 
radio sources with intermediate or low flux densities are an important 
ingredient to modern astrophysics. We have therefore embarked on a project 
to obtain flux densities for the B3-VLA sample (Vigotti et al. 1989) over 
a frequency range as wide as possible. The aim is to study the spectral 
properties of a complete sample of radio sources.

The B3-VLA sample is composed of sources that are roughly equally distributed 
in five flux density intervals, i.e. 50 times fainter than the 3C survey
(Bennett 1962). The sample now contains 1049 radio sources, instead of 1050
listed in the previous papers dealing with the B3-VLA sample: the source 
2302+396 which was already indicated as a possible spurious source close to 
a grating ring in the B3 catalogue (Ficarra et al. 1985) was deleted from
the list since it was recognized as a CLEAN artifact. In fact it could not 
be found neither in the WENSS nor in the NVSS catalogue.

This paper is the second of a series describing the multifrequency properties 
of the B3-VLA sample. In the first paper we presented the radio continuum data 
at 10.6~GHz obtained with the Effelsberg radio telescope (Gregorini et al.
1998, 
hereafter Paper I). We detected 99$\%$ of the radio sources, with a typical
flux density error of about 1~mJy for the fainter ones.

Here we present the spectral database of the whole sample consisting of flux
densities at 151~MHz, 327~MHz, 408~MHz, 1.4~GHz, 4.85~GHz, and 10.6~GHz.
Additional observations were performed for 478 sources at 4.85~GHz, which were 
necessary to complete the information at this frequency and to measure also at
4.85 GHz the polarization detected at 10.6 GHz. 

Sect.~2 describes the observations and data reduction at 4.85~GHz. In Sect.~3 
we present the database, with an accurate description of the method used to 
obtain the flux densities and the errors at each frequency. In Sect.~4 the
data table is presented with a discussion of the data quality.

\section{Observations at 4.85 GHz}

The observations reported here have been carried out between July 1994 and 
March 1999. Until August 1995 the old $\lambda$6-cm correlation receiver 
system was employed. This system had two feeds in the secondary focus of the 
100-m telescope. The right-hand circular polarization outputs from each feed 
(obtained after the polarizers in the wave\-guides) were correlated via a 3-dB 
hybrid to yield a differential total-power signal of the two feeds. This 
double--beam ensured minimal atmospheric disturbance to the signal. 
Amplification in the first stage was achieved with cooled FET's. The main 
horn was connected to an IF-polarimeter to deliver the Stokes U and Q 
parameters for full linear polarization information. The system operated at 
a centre frequency of 4.75~GHz, with a bandwidth of 500~MHz. The receiver 
system temperature was $\sim$70~K on the sky (zenith, clear sky). 

In August 1995 this receivers were replaced by two stable total-power systems, 
with HEMT amplifiers in the first stage. Here the differential signal is 
retrieved by subtracting the calibrated signals in the computer. Each of the 
two total-power receivers is connected to an IF-polarimeter. This system 
operates at 4.85~GHz, the bandwidth is 500~MHz. The receiver system 
temperature has been greatly improved to 30~K on the sky (zenith, 
clear sky).

The half-power beam width was 147\arcsec~for the old and 143\arcsec~for the 
new receiver system and the beam throw was 8\farcm2 in both cases.
The sources were observed by cross-scanning the 
telescope in right ascension and declination, with a scan length of 15\arcmin. 
The scanning speed was 30\arcmin/min., and the total number of scans was 
adjusted to the expected flux density of each source. Sources with angular 
extents significantly exceeding the beam size or exhibiting significant 
confusion in the cross-scans were mapped in the double-beam mode and 
subsequently restored to the equivalent single-beam images using the 
restoration algorithm of Emerson et al. (1979). The scan separation was 
1\arcmin, and the map sizes adjusted such as to account for the source 
size and the beam separation. The total number of sources mapped this way 
is 6. 

Telescope pointing, focussing and polarimeter adjustments were regularly 
checked by cross-scanning the point sources NGC\,7027, 3C\,48, 3C\,84, 
3C\,138, 3C\,147, 3C\,196, 3C\,286 and 3C\,295. The latter two sources 
served also as flux density calibrators. 

\section{Database}

In Tab.~1 we present the information available for the B3--VLA sample.
Cols.~1 and 2 list the frequency and reference to the relevant paper,
Col.~3 gives the percentage of sources for which the data is available.

\begin{table}
\caption[]{Measured flux densities of B3-VLA sources}
\begin{flushleft}
\begin{tabular}{rlr}
\hline\noalign{\smallskip}
  Frequency  & Reference & $\%$ \\
\noalign{\smallskip}
\hline\noalign{\smallskip}
   151~MHz & Hales et al. 1988 & 89 \\
   327~MHz & Rengelink et al. 1997 & 100 \\
   408~MHz & Ficarra et al.  1985 & 100 \\
   1.4~GHz & Condon et al.  1998 & 100 \\
   4.85~GHz & present paper  & 100 \\
            & Kulkarni et al. 1990 & 60 \\
            & Gregory et al. 1996 & 83  \\
  10.6~GHz  & Gregorini et al. 1998 & 99 \\
\noalign{\smallskip}
\hline
\end{tabular}
\end{flushleft}
\end{table}

\subsection{151~MHz Data }

These flux densities were obtained by cross-correlating the 6C survey (Hales 
et al. 1988) with the B3-VLA sample. The search radius used was 100\arcsec, 
which corresponds to a combined 3-$\sigma$ error for the fainter sources.
We do not 
expect any chance coincidences, owing to the low source density at 151~MHz 
(4.1 sources per square degree). The values quoted in Tab.~2 are the peak 
flux densities for sources with an angular extent $<$~100\arcsec~(extents
taken from Vigotti et al. 1989), and the 
integrated ones (listed in the 6C, Hales et al. 1988) for larger sources.
The error in the same table is computed 
as a constant term of 40 mJy, plus a 5\% contribution due to the uncertainty 
of the flux density scale.

Since these data are on the flux scale of Roger et al. (1973, RBC), we used the
spectral indices reported by these authors to calculate the flux density of 
their calibrator sources at 178 MHz. In this way we could compare the scale
of Roger et al. (1973) with the one of Kellermann et al. (1969, KPW). The ratio
between these two flux density scales is KPW/RBC = 0.96. Baars et al. (1977,
BGPW) report a ratio of BGPW/KPW = 1.051. Thus, the ratio BGPW/RBC turns out
to be 1.008; therefore no correction was applied at 151 MHz. 
\subsection{327~MHz Data}

For sources with an angular extent $<$~50\arcsec\, we cross-correlated the 
B3--VLA positions with the WENSS source list (Rengelink et al. 1997) using a 
window of 11\arcsec\, in right ascension and 22\arcsec\, in declination. For 
the more extended ones we used a window of 40\arcsec\, in right ascension and 
80\arcsec\, in declination. The total area searched was 0.03 square degrees. 
The WENSS source density is about 21.3 per square degree so that the 
contamination by chance coincidences is negligible. For the flux density
errors we used the formula 
given by Rengelink et al. (1997), with a noise contribution of 4.5~mJy (which 
is the average value in the B3--VLA area), plus 4\% due to the calibration 
uncertainty $\Delta_{\rm cal}$ . 

Sources with a complex structure (as marked in the WENSS catalogue;
Rengelink et al. 1997) 
were inspected directly on the WENSS maps, and their flux densities computed 
with the AIPS task TVSTAT. For these sources the errors $\Delta$S were computed 
as follows:

$$\Delta \rm S = \sqrt {(\Delta_{\rm cal} \cdot {\rm S})^2 +\sigma_l^2\cdot
 \frac{A_s}{A_b}}$$.

\noindent 
Here $\sigma_{\rm l}$ is the local noise in the map, A$_{\rm s}$ the area
covered 
by the radio source, and A$_{\rm b}$ is the beam area. The flux densities in 
the WENSS survey are on the scale of Baars et al. (1977).

\subsection{408~MHz Data }

The flux densities were taken from the B3 survey, except for extended sources
for which an integrated flux density was used (Vigotti et al. 1989).
For the computation 
of the errors we used 35~mJy as the constant term and 3$\%$ for the term 
proportional to the source flux density (Ficarra et al. 1985).
The flux density scale 
of these data is based on 3C123, and agrees with the scale of Baars et al. 
(1977) to within 2$\%$. Therefore, no correction was applied. 

\subsection{1.4~GHz Data }

The flux densities were computed from the maps of the NRAO VLA SKY Survey 
(NVSS, Condon et al. 1998), centred on the B3-VLA positions using an automatic 
two-component Gaussian fit algorithm similar to the AIPS task JMFIT. For the 
unresolved sources the difference between our flux density and that listed in 
the NVSS catalogue is negligible ($<$~2\%). The errors were calculated with 
the formula of Condon et al. (1998), where the noise and confusion term is 
0.45~mJy/beam and the calibration uncertainty is 3\%.

For the extended and complex sources the flux densities were computed using
the AIPS task TVSTAT. Their errors were computed as above (Sect.~3.2).
The flux densities are on the scale of Baars et al. (1977).
\subsection{4.85~GHz Data }

All sources not available in the literature (Kulkarni et al. 1990, Gregory et 
al. 1996) have been observed as described in Sect.~2. The flux densities of 
Kulkarni et al. 1990, and those observed by us before August 1995 were shifted 
from 4.75~GHz to 4.85~GHz using the spectral index of the radio source. For 
the flux densities presented in this paper we adopted 1.0~mJy as the noise 
contribution, and 2\% as the contribution proportional to the flux density. 
Another 0.45~mJy is added to account for source confusion (Reich 1993).
For the data of 
Kulkarni et al. (1990) the errors are 2~mJy and 2\%, respectively. The errors 
of the flux densities taken from Gregory et al. (1996) are listed in the GB6 
catalogue. In 2 cases, 1412+397 and 2341+396B, the sources could not be 
separated from a closeby confusing source.
We used the flux densities from our measurements (45.8\%). In cases where
those were not available the flux densities reported by Kulkarni et al. (1990;
42.4\%) or Gregory et al. (1996; 11.8\%) were taken.

The GB6 maps of the sources with extension larger than 70\arcsec were 
downloaded using {\it SkyView}. In addition, the most extended ones (0136+396,
0157+405A, 0248+467, 0703+426A, 1141+374, and 1309+412A) were mapped in 
Effelsberg. In all cases the flux densities were determined with AIPS task
TVSTAT and the errors were calculated as described above (Sect.~3.2).
The flux densities are on the scale of Baars et al. (1977).

\subsection{10.6~GHz Data } 

In Tab.~2 we list the integrated flux densities as well as the errors computed 
using the formula presented in Paper I. Here, the noise term is 0.8~mJy, 
confusion contributes 0.08~mJy, and the term proportional to the flux is 2\%. 
The flux densities are on the scale of Baars et al. (1977).

\section{Discussion}

\begin{figure}
 \resizebox{\hsize}{!}{\includegraphics{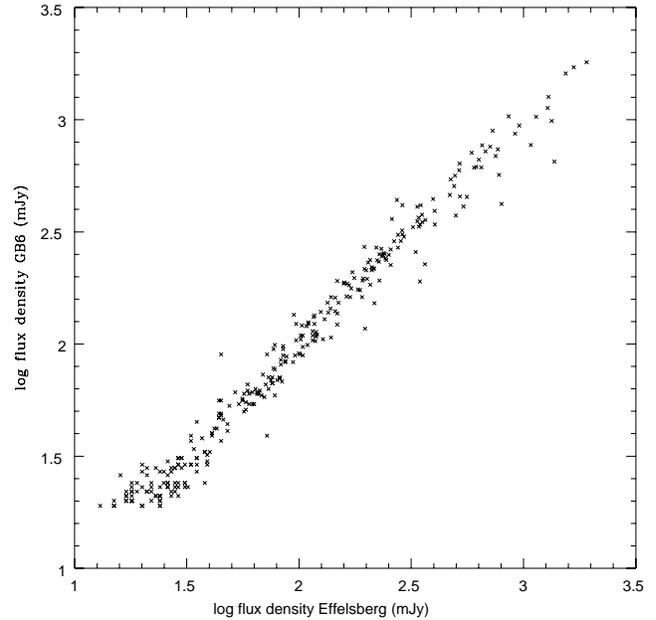}}
 \caption{Flux densities of the GB6 survey versus our measurement}
\end{figure}
\begin{figure}
 \resizebox{\hsize}{!}{\includegraphics{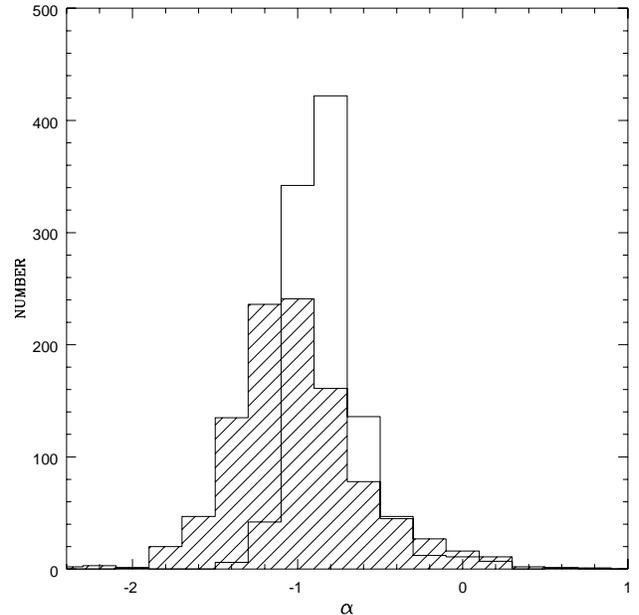}}
 \caption{Histogram of low ($\alpha_l$)- and high-frequency ($\alpha_h$)
spectral indices. The blank area corresponds to $\alpha_l$, the hashed one
to $\alpha_h$.}
\end{figure}
\begin{figure}
 \resizebox{\hsize}{!}{\includegraphics{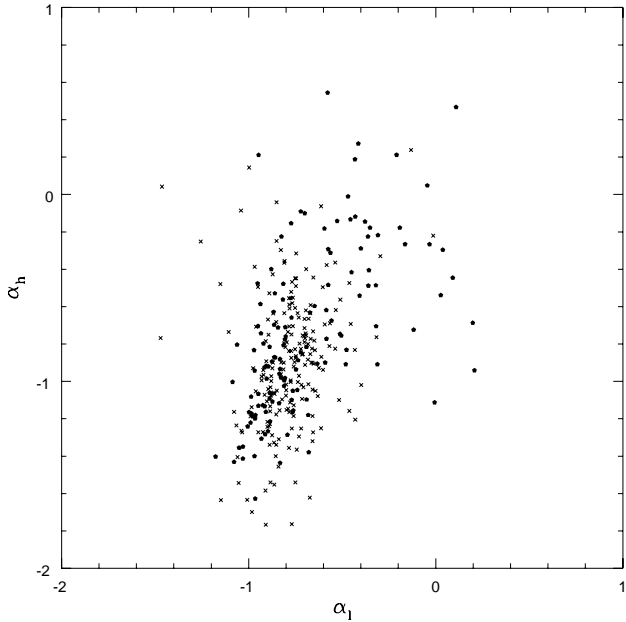}}
 \caption{Colour-colour diagram of B3VLA sources. Galaxies are symbolized 
by crosses, quasars by dots}
\end{figure}
Table~2 presents the whole database. Col.~1 lists the B3-VLA name, Cols.~2
and 3 the radio centroid (equinox J2000.0) from Vigotti et al. (1989;
computed as the geometric mean of the source components). The following 12
columns list the flux densities and errors at 151~MHz, 327~MHz, 408~MHz,
1.4~GHz, 4.85~GHz and 10.6~GHz, respectively (all in mJy). The last
column contains the sources' optical identifications, abbreviated as follows:
g: radio galaxy identified on the POSS-I, most of which are z$\le 0.5$; 
G: far radio galaxy with measured redshift (0.5$\le{\rm z}\le$ 3.5); Q: 
spectroscopically confirmed quasar; b: blue objects (i.e. non-confirmed
quasars); BL: BL Lac; F: featureless spectrum; a blank means `empty field',
i.e. it lacks any optical counterpart down to the 
POSS-I limit (more that 90\% are distant radio galaxies,
the remaining ones being quasars with magnitudes fainter than the POSS-I).

For 19 sources the 408~MHz data are not reported. In 15 cases the flux 
density is affected by a nearby strong source. In four cases the B3-VLA
sources were not resolved by the 408~MHz beam.

In order to complete the spectral database we observed 164 sources at 4.85~GHz 
whose flux densities were not available in the catalogues listed in Tab.~1;
314 sources with 
detected polarization at 10.6~GHz were re-observed at 4.85 GHz for future
polarization studies. An analysis of the polarization data will be published 
in a forthcoming paper. In Fig.~1 we show the plot of our measurements versus
the GB6 flux densities. Intrinsic source variability is likely to increase
the scatter of the plot.

For each source we computed two spectral indices: a low-frequency index 
$\alpha_l$ (0.3 -- 1.4~GHz) and a high-frequency one, $\alpha_h$ (4.8 -- 
10.6~GHz). Fig.~2 shows the histograms of $\alpha_l$ and $\alpha_h$ (shaded) 
of 1034 sources, for which four flux densities are available. The resulting 
median values for the two distributions are
$\langle\alpha_l\rangle = -0.853$ 
and $\langle\alpha_h\rangle = -1.053$ (S $\propto \nu^{\alpha}$). 

In Fig.~3 we show a radio colour-colour diagram illustrating the different 
population areas of radio galaxies and quasars. As already evident in Fig.~2, 
$\alpha_h$ covers a wider range of values (dispersion 0.40) than $\alpha_l$ 
(dispersion 0.23). This is to be expected if spectral steepening due to 
synchrotron and inverse Compton energy losses is important: it changes 
$\alpha_h$ first, before the sources have aged sufficiently such as to affect 
$\alpha_l$ as well. The corresponding evolutionary track in the $\alpha_l$ -- 
$\alpha_h$ diagramme is that populated by the radio galaxies in Fig.~3: if 
these sources commence their lives with flat (injection) spectra, they should 
gradually move downward at a faster rate than leftward. Also evident in Fig.~3 
is that radio galaxies (crosses) have on average steeper high-frequency 
spectra than quasars; in particular, the radio galaxies dominate the lowest
part of the diagramme.

Some sources (essentially quasars) exhibit extreme values, especially those 
with flat $\alpha_l$ and/or flat $\alpha_h$ (populating the upper and 
right-hand portion of the plot). These may possess self-absorbed components 
that become visible in different frequency regimes, depending on their 
optical thickness. A thorough analysis and interpretation of our results 
will be presented in a forthcoming paper. 

\begin{acknowledgements}
We thank Helge Rottmann for his help during the Effelsberg observations.
We are grateful to Dr. Heinz Andernach whose comments on the manuscript 
helped to improve the paper significantly.
Part of this work was supported by the Deutsche Forschungsgemeinschaft, grant 
KL533/4-2, and by the European Commission, TMR Programme, Research Network 
Contract ERBFMRXCT97-0034 ``CERES''. We thank the Italian Ministry for
University and Scientic Research (MURST) for partial financial support
(grant Cofin98-02-32). We acknowledge the use NASA's {\it SkyView} facility
(http://skyview.gsfc.nasa.gov) located at NASA Goddard Space Flight Center.
\end{acknowledgements}

\tabcolsep0.17cm
\begin{table*}
\caption[]{B3\,VLA flux densities}
\begin{flushleft}
\begin{tiny}
\begin{tabular}{|l|cc|rrrrrrrrrrrr|c|}
\hline
%B3 name  &RA(J2000)&DEC(J2000)&\multicolumn{1}{c}{S$_{151}$}&\multicolumn{1}{c}{$\Delta$S$_{151}$}&\multicolumn{1}{c}{S$_{327}$}&\multicolumn{1}{c}{$\Delta$S$_{327}$}&\multicolumn{1}{c}{S$_{408}$}&\multicolumn{1}{c}{$\Delta$S$_{408}$}&\multicolumn{1}{c}{S$_{1400}$}&\multicolumn{1}{c}{$\Delta$S$_{1400}$}&\multicolumn{1}{c}{S$_{4850}$}&\multicolumn{1}{c}{$\Delta$S$_{4850}$}&\multicolumn{1}{c}{S$_{10550}$}&\multicolumn{1}{c}{$\Delta$S$_{10550}$}& Id.\\
B3 name  &RA(J2000)&DEC(J2000)&{S$_{151}$}&{$\Delta$S$_{151}$}&{S$_{327}$}&{$\Delta$S$_{327}$}&{S$_{408}$}&{$\Delta$S$_{408}$}&{S$_{1400}$}&{$\Delta$S$_{1400}$}&{S$_{4850}$}&{$\Delta$S$_{4850}$}&{c}{S$_{10550}$}&{$\Delta$S$_{10550}$}& Id.\\
         & [$^{\rm h}$ ~ $^{\rm m}$ ~ $^{\rm s}$] & [\degr$\;\;\;$\arcmin$\;\;\;$\arcsec] & [mJy] & [mJy] & [mJy] & [mJy]& [mJy] & [mJy] &[mJy]&
[mJy]&[mJy]&[mJy]&[mJy]&[mJy]& \\  
\hline
0000+394 &00 02 42.4&39 46 34&   360&   44&   270&   12&   190&  31&    93&   5&    36&   7&    18&   1&   \\
0000+399 &00 03 32.0&40 13 26&      &     &   140&    7&   150&  30&    53&   2&    17&   1&    13&   1&   \\
0001+395 &00 04 15.9&39 49 52&   530&   48&   200&    9&   130&  30&    33&   1&     6&   1&     2&   1&   \\
0001+398 &00 04 19.5&40 06 09&   390&   45&   220&   10&   220&  31&    76&   2&    30&   4&    11&   1&g  \\
0003+380 &00 05 57.1&38 20 15&   790&   56&   580&   24&   490&  33&   568&  17&   778&  16&   655&  13&g  \\
0003+387 &00 06 20.8&39 00 27&  2910&  151&  1750&   70&  1360&  51&   488&  15&   107&   2&    30&   1&G  \\
\hline    
\end{tabular}
\end{tiny}
\end{flushleft}
\end{table*}
\clearpage
\setcounter{table}{1}
\tabcolsep0.17cm
\begin{table*}
\caption[]{B3\,VLA flux densities (cont'd)}
\begin{flushleft}
\begin{tiny}
\begin{tabular}{|l|cc|rrrrrrrrrrrr|c|}
\hline
B3 name  &RA(J2000)&DEC(J2000)&S$_{151}$&$\Delta$S$_{151}$&S$_{327}$&$\Delta$S$_{327}$&S$_{408}$&$\Delta$S$_{408}$&S$_{1400}$&$\Delta$S$_{1400}$&S$_{4850}$&$\Delta$S$_{4850}$&S$_{10550}$&$\Delta$S$_{10550}$&Id.\\
         & [$^{\rm h}$ ~ $^{\rm m}$ ~ $^{\rm s}$] & [\degr$\;\;\;$\arcmin$\;\;\;$\arcsec] & [mJy] & [mJy] & [mJy] & [mJy]& [mJy] & [mJy] &[mJy]&
[mJy]&[mJy]&[mJy]&[mJy]&[mJy]& \\  
\hline
0109+415 &01 12 08.3&41 46 60&  2350&  124&  1390&   56&  1140&  45&   541&  16&   197&   4&   107&   2&g  \\
0109+416B&01 12 15.4&41 55 13&  4170&  212&  2620&  105&  1960&  66&   856&  26&   247&  27&    60&   2&g  \\
0110+386 &01 12 57.4&38 53 17&  1200&   72&   580&   24&   490&  33&   174&   5&    56&   2&    22&   1&   \\
0110+395 &01 13 37.6&39 50 33&   860&   59&   390&   16&   300&  31&   109&   3&    48&   5&    39&   1&   \\
0110+398 &01 13 28.7&40 10 28&   300&   43&   190&    9&   130&  30&    46&   1&    14&   1&     6&   1&   \\
0110+401 &01 13 17.7&40 26 12&  1820&   99&  1290&   52&  1080&  44&   551&  17&   233&   5&   144&   3&Q  \\
\hline    
\end{tabular}
\end{tiny}
\end{flushleft}
\end{table*}
\clearpage
\setcounter{table}{1}
\tabcolsep0.17cm
\begin{table*}
\caption[]{B3\,VLA flux densities (cont'd)}
\begin{flushleft}
\begin{tiny}
\begin{tabular}{|l|cc|rrrrrrrrrrrr|c|}
\hline
B3 name  &RA(J2000)&DEC(J2000)&S$_{151}$&$\Delta$S$_{151}$&S$_{327}$&$\Delta$S$_{327}$&S$_{408}$&$\Delta$S$_{408}$&S$_{1400}$&$\Delta$S$_{1400}$&S$_{4850}$&$\Delta$S$_{4850}$&S$_{10550}$&$\Delta$S$_{10550}$&Id.\\
         & [$^{\rm h}$ ~ $^{\rm m}$ ~ $^{\rm s}$] & [\degr$\;\;\;$\arcmin$\;\;\;$\arcsec] & [mJy] & [mJy] & [mJy] & [mJy]& [mJy] & [mJy] &[mJy]&
[mJy]&[mJy]&[mJy]&[mJy]&[mJy]& \\  
\hline
0220+427A&02 23 16.5&42 59 39& 20950& 1048& 21410&  856& 14830& 446&  6214& 186&  3060& 306&  1806&  36&g  \\
0221+383 &02 24 16.9&38 32 08&  1080&   67&   570&   23&   430&  33&   153&   5&    55&  11&    29&   1&   \\
0221+393 &02 24 43.2&39 32 47&   640&   51&   270&   12&   210&  31&    53&   4&     7&   2&     8&   1&   \\
0221+396 &02 24 50.7&39 55 16&      &     &   180&    9&   220&  31&    58&   2&    32&   4&    33&   1&   \\
0222+403 &02 25 44.8&40 31 35&  4050&  206&  2490&  100&  2190&  72&   941&  28&   351&  36&   182&   4&g  \\
0222+422A&02 25 32.2&42 29 41&  2000&  108&  1130&   45&   810&  39&   262&   8&    66&   2&    23&   1&   \\
\hline    
\end{tabular}
\end{tiny}
\end{flushleft}
\end{table*}
\clearpage
\setcounter{table}{1}
\tabcolsep0.17cm
\begin{table*}
\caption[]{B3\,VLA flux densities (cont'd)}
\begin{flushleft}
\begin{tiny}
\begin{tabular}{|l|cc|rrrrrrrrrrrr|c|}
\hline
B3 name  &RA(J2000)&DEC(J2000)&S$_{151}$&$\Delta$S$_{151}$&S$_{327}$&$\Delta$S$_{327}$&S$_{408}$&$\Delta$S$_{408}$&S$_{1400}$&$\Delta$S$_{1400}$&S$_{4850}$&$\Delta$S$_{4850}$&S$_{10550}$&$\Delta$S$_{10550}$&Id.\\
         & [$^{\rm h}$ ~ $^{\rm m}$ ~ $^{\rm s}$] & [\degr$\;\;\;$\arcmin$\;\;\;$\arcsec] & [mJy] & [mJy] & [mJy] & [mJy]& [mJy] & [mJy] &[mJy]&
[mJy]&[mJy]&[mJy]&[mJy]&[mJy]& \\  
\hline
0726+431 &07 29 48.3&43 01 19&  3010&  156&  1780&   71&  1290&  49&   425&  13&   126&   3&    50&   1&Q  \\
0727+401 &07 30 35.8&40 01 23&   290&   43&   390&   16&   400&  32&   451&  14&   144&   3&    49&   1&   \\
0728+389 &07 31 57.9&38 51 26&   340&   43&   280&   12&   200&  31&   111&   3&    35&   5&    22&   1&   \\
0728+395 &07 31 55.3&39 23 35&   640&   51&   400&   17&   300&  31&    92&   3&    22&   1&    10&   1&   \\
0729+391 &07 33 20.9&39 05 06&   290&   43&   280&   12&   260&  31&   155&   5&   100&   9&    66&   2&Q  \\
0729+395 &07 33 00.9&39 25 06&   600&   50&   380&   16&   280&  31&   105&   3&    35&   5&    10&   1&g  \\
\hline    
\end{tabular}
\end{tiny}
\end{flushleft}
\end{table*}
\clearpage
\setcounter{table}{1}
\tabcolsep0.17cm
\begin{table*}
\caption[]{B3\,VLA flux densities (cont'd)}
\begin{flushleft}
\begin{tiny}
\begin{tabular}{|l|cc|rrrrrrrrrrrr|c|}
\hline
B3 name  &RA(J2000)&DEC(J2000)&S$_{151}$&$\Delta$S$_{151}$&S$_{327}$&$\Delta$S$_{327}$&S$_{408}$&$\Delta$S$_{408}$&S$_{1400}$&$\Delta$S$_{1400}$&S$_{4850}$&$\Delta$S$_{4850}$&S$_{10550}$&$\Delta$S$_{10550}$&Id.\\
         & [$^{\rm h}$ ~ $^{\rm m}$ ~ $^{\rm s}$] & [\degr$\;\;\;$\arcmin$\;\;\;$\arcsec] & [mJy] & [mJy] & [mJy] & [mJy]& [mJy] & [mJy] &[mJy]&
[mJy]&[mJy]&[mJy]&[mJy]&[mJy]& \\  
\hline
0827+458 &08 30 35.8&45 43 30&  6260&  316&  3050&  122&  2360&  77&   619&  19&   134&   3&    48&   1&   \\
0828+381 &08 31 51.0&37 56 32&   960&   62&   560&   23&   460&  33&   156&   5&    40&   1&    19&   1&g  \\
0829+395 &08 33 15.4&39 21 18&  1500&   85&   850&   34&   630&  35&   226&   7&    64&   2&    30&   1&   \\
0829+425 &08 32 48.5&42 24 59&  1500&   85&  1030&   41&   820&  39&   458&  14&   229&   5&   135&   3&Q  \\
0831+393 &08 34 55.0&39 10 52&   330&   43&   240&   11&   210&  31&    76&   2&    27&   4&    15&   1&   \\
0831+399 &08 34 33.8&39 44 53&   690&   53&   360&   15&   250&  31&   103&   3&    27&   4&    17&   1&   \\
\hline    
\end{tabular}
\end{tiny}
\end{flushleft}
\end{table*}
\clearpage
\setcounter{table}{1}
\tabcolsep0.17cm
\begin{table*}
\caption[]{B3\,VLA flux densities (cont'd)}
\begin{flushleft}
\begin{tiny}
\begin{tabular}{|l|cc|rrrrrrrrrrrr|c|}
\hline
B3 name  &RA(J2000)&DEC(J2000)&S$_{151}$&$\Delta$S$_{151}$&S$_{327}$&$\Delta$S$_{327}$&S$_{408}$&$\Delta$S$_{408}$&S$_{1400}$&$\Delta$S$_{1400}$&S$_{4850}$&$\Delta$S$_{4850}$&S$_{10550}$&$\Delta$S$_{10550}$&Id.\\
         & [$^{\rm h}$ ~ $^{\rm m}$ ~ $^{\rm s}$] & [\degr$\;\;\;$\arcmin$\;\;\;$\arcsec] & [mJy] & [mJy] & [mJy] & [mJy]& [mJy] & [mJy] &[mJy]&
[mJy]&[mJy]&[mJy]&[mJy]&[mJy]& \\  
\hline
0944+397 &09 47 49.1&39 33 11&   660&   52&   350&   15&   250&  31&    72&   2&    18&   1&     5&   1&   \\
0945+408 &09 48 55.4&40 39 44&  3240&  167&  2700&  108&  2450&  79&  1606&  48&  1910&  38&  1392&  28&Q  \\
0945+419 &09 48 30.2&41 41 43&  5420&  274&  2960&  118&  2210&  73&   613&  18&   138&   3&    41&   1&   \\
0947+405 &09 50 28.2&40 17 56&  1030&   65&   530&   22&   440&  33&   127&   4&    35&   1&    16&   1&   \\
0947+424 &09 50 18.5&42 12 58&  3780&  193&  2020&   81&  1480&  54&   532&  16&   147&   3&    60&   2&   \\
0948+390 &09 52 01.3&38 50 32&   800&   57&   500&   21&   400&  32&   144&   4&    46&   1&    15&   1&g  \\
\hline    
\end{tabular}
\end{tiny}
\end{flushleft}
\end{table*}
\clearpage
\setcounter{table}{1}
\tabcolsep0.17cm
\begin{table*}
\caption[]{B3\,VLA flux densities (cont'd)}
\begin{flushleft}
\begin{tiny}
\begin{tabular}{|l|cc|rrrrrrrrrrrr|c|}
\hline
B3 name  &RA(J2000)&DEC(J2000)&S$_{151}$&$\Delta$S$_{151}$&S$_{327}$&$\Delta$S$_{327}$&S$_{408}$&$\Delta$S$_{408}$&S$_{1400}$&$\Delta$S$_{1400}$&S$_{4850}$&$\Delta$S$_{4850}$&S$_{10550}$&$\Delta$S$_{10550}$&Id.\\
         & [$^{\rm h}$ ~ $^{\rm m}$ ~ $^{\rm s}$] & [\degr$\;\;\;$\arcmin$\;\;\;$\arcsec] & [mJy] & [mJy] & [mJy] & [mJy]& [mJy] & [mJy] &[mJy]&
[mJy]&[mJy]&[mJy]&[mJy]&[mJy]& \\  
\hline
1103+393 &11 06 08.2&39 06 04&  1170&   71&   660&   27&   520&  34&   190&   6&    62&   2&    28&   1&   \\
1104+390 &11 07 40.3&38 49 03&   370&   44&   230&   10&   200&  31&    71&   2&    25&   1&    15&   1&   \\
1104+397 &11 07 01.0&39 29 13&   350&   44&   160&    8&   120&  30&    30&   2&     5&   1&   $<$4&    &   \\
1105+390 &11 08 22.5&38 45 52&   470&   46&   280&   12&   230&  31&    91&   3&    29&   5&    19&   1&g  \\
1105+392 &11 08 37.9&38 58 39&  4990&  253&  2900&  116&  2310&  76&   898&  27&   276&   6&   128&   3&Q  \\
1106+380 &11 09 28.8&37 44 32&  1920&  104&  1370&   55&  1290&  49&  1226&  37&   769&  15&   502&  10&G  \\
\hline    
\end{tabular}
\end{tiny}
\end{flushleft}
\end{table*}
\clearpage
\setcounter{table}{1}
\tabcolsep0.17cm
\begin{table*}
\caption[]{B3\,VLA flux densities (cont'd)}
\begin{flushleft}
\begin{tiny}
\begin{tabular}{|l|cc|rrrrrrrrrrrr|c|}
\hline
B3 name  &RA(J2000)&DEC(J2000)&S$_{151}$&$\Delta$S$_{151}$&S$_{327}$&$\Delta$S$_{327}$&S$_{408}$&$\Delta$S$_{408}$&S$_{1400}$&$\Delta$S$_{1400}$&S$_{4850}$&$\Delta$S$_{4850}$&S$_{10550}$&$\Delta$S$_{10550}$&Id.\\
         & [$^{\rm h}$ ~ $^{\rm m}$ ~ $^{\rm s}$] & [\degr$\;\;\;$\arcmin$\;\;\;$\arcsec] & [mJy] & [mJy] & [mJy] & [mJy]& [mJy] & [mJy] &[mJy]&
[mJy]&[mJy]&[mJy]&[mJy]&[mJy]& \\  
\hline
1217+427 &12 19 53.9&42 29 50&  1820&   99&  1130&   45&   870&  40&   305&   9&    88&   2&    37&   1&   \\
1218+395 &12 21 11.5&39 18 46&  1010&   64&   500&   21&   330&  32&   104&   3&    27&   1&    11&   1&   \\
1218+398 &12 20 53.8&39 37 12&   230&   42&   110&    6&   110&  30&    31&   1&     9&   1&     4&   1&   \\
1218+421 &12 21 12.6&41 52 18&  2000&  108&  1310&   53&  1080&  44&   501&  15&   193&   4&   101&   2&g  \\
1219+382 &12 22 11.4&37 59 14&  1510&   85&   830&   34&   580&  35&   175&   5&    30&   1&     9&   1&   \\
1220+393 &12 23 22.5&39 04 23&   740&   54&   410&   17&   290&  31&   103&   3&    27&   1&    12&   1&   \\
\hline    
\end{tabular}
\end{tiny}
\end{flushleft}
\end{table*}
\clearpage
\setcounter{table}{1}
\tabcolsep0.17cm
\begin{table*}
\caption[]{B3\,VLA flux densities (cont'd)}
\begin{flushleft}
\begin{tiny}
\begin{tabular}{|l|cc|rrrrrrrrrrrr|c|}
\hline
B3 name  &RA(J2000)&DEC(J2000)&S$_{151}$&$\Delta$S$_{151}$&S$_{327}$&$\Delta$S$_{327}$&S$_{408}$&$\Delta$S$_{408}$&S$_{1400}$&$\Delta$S$_{1400}$&S$_{4850}$&$\Delta$S$_{4850}$&S$_{10550}$&$\Delta$S$_{10550}$&Id.\\
         & [$^{\rm h}$ ~ $^{\rm m}$ ~ $^{\rm s}$] & [\degr$\;\;\;$\arcmin$\;\;\;$\arcsec] & [mJy] & [mJy] & [mJy] & [mJy]& [mJy] & [mJy] &[mJy]&
[mJy]&[mJy]&[mJy]&[mJy]&[mJy]& \\  
\hline
1333+392 &13 35 50.6&38 59 55&   920&   61&   450&   19&   370&  32&   120&   4&    38&   1&    15&   1&   \\
1333+412 &13 35 19.9&41 00 04&  7460&  375&  3840&  154&  2980&  94&   852&  26&   256&   5&    95&   2&g  \\
1334+417 &13 36 26.4&41 31 12&  1440&   82&  1050&   42&   890&  40&   449&  13&   171&   4&    62&   2&G  \\
1336+391A&13 38 49.7&38 51 12& 20270& 1014& 12510&  500& 10000& 301&  3377& 101&   958&  19&   363&   7&g  \\
1336+393 &13 38 15.3&39 05 57&   980&   63&   460&   19&   400&  32&   113&   3&    45&   1&    16&   1&   \\
1336+396C&13 39 08.7&39 26 22&  2840&  148&  1540&   62&  1180&  46&   415&  12&   135&   3&    58&   2&   \\
\hline    
\end{tabular}
\end{tiny}
\end{flushleft}
\end{table*}
\clearpage
\setcounter{table}{1}
\tabcolsep0.17cm
\begin{table*}
\caption[]{B3\,VLA flux densities (cont'd)}
\begin{flushleft}
\begin{tiny}
\begin{tabular}{|l|cc|rrrrrrrrrrrr|c|}
\hline
B3 name  &RA(J2000)&DEC(J2000)&S$_{151}$&$\Delta$S$_{151}$&S$_{327}$&$\Delta$S$_{327}$&S$_{408}$&$\Delta$S$_{408}$&S$_{1400}$&$\Delta$S$_{1400}$&S$_{4850}$&$\Delta$S$_{4850}$&S$_{10550}$&$\Delta$S$_{10550}$&Id.\\
         & [$^{\rm h}$ ~ $^{\rm m}$ ~ $^{\rm s}$] & [\degr$\;\;\;$\arcmin$\;\;\;$\arcsec] & [mJy] & [mJy] & [mJy] & [mJy]& [mJy] & [mJy] &[mJy]&
[mJy]&[mJy]&[mJy]&[mJy]&[mJy]& \\  
\hline
1446+399 &14 48 07.4&39 44 20&   750&   55&   440&   18&      &    &   103&   5&    10&   2&    11&   1&   \\
1446+440 &14 48 32.0&43 52 35&  3910&  200&  2050&   82&  1420&  52&   543&  16&   154&   3&    70&   2&   \\
1447+380 &14 49 17.0&37 48 42&  1260&   75&   660&   27&   420&  33&   114&   3&    27&   1&    10&   1&   \\
1447+400 &14 49 29.6&39 48 23&  2310&  122&  1330&   53&   970&  42&   322&  10&   102&   2&    50&   1&   \\
1447+402 &14 49 02.3&40 00 21&  2300&  122&  1730&   73&  1120&  45&   501&  15&   178&   4&   104&   2&g  \\
1449+380 &14 50 59.8&37 51 31&  1070&   67&   690&   28&   520&  34&   193&   6&    51&   1&    17&   1&   \\
\hline    
\end{tabular}
\end{tiny}
\end{flushleft}
\end{table*}
\clearpage


\begin{thebibliography}{}
\bibitem[Baars et al. 1977]{Ba:77}
Baars J.W.M., Genzel R., Pauliny-Toth I.I.K., Witzel A.,
1977, A\&A 61, 99
\bibitem[Bennett 1962]{Be:62}
Bennett A. S., 1962, Mem.R.astr.Soc., 68, 163
\bibitem[Condon et al.]{Co:98}
Condon J.J., Cotton W.D., Greisen E.W., Yin Q.F., Perley R.A., Taylor G.B., 
Broderick J.J., 1998, AJ 115, 1693
\bibitem[Emerson et al. 1979]{Em:79}
Emerson D.T., Klein U., Haslam C.G.T., 1979, A\&A 76, 92
\bibitem[Ficarra et al. 1985]{Fi:90}
Ficarra A., Grueff G., Tomassetti G., 1985, A\&AS 59, 255
\bibitem[Gregory et al. 1996]{Gy:92}
Gregory P.C., Scott W.K., Douglas K., Condon J.J., 1996, ApJS 103, 427
\bibitem[Gregorini et al. 1992]{Gr:92}
Gregorini L., Vigotti M., Mack K.-H., Z\"onnchen J., Klein U., 1998, 
A\&AS 133, 129 (Paper I)
\bibitem[Hales et al. 1988]{Ha:88}
Hales S.E.G., Baldwin J.E., Warner P.J., 1988, MNRAS 234, 919
\bibitem[Kellermann et al. 1969]{Ke:69}
Kellermann K.L., Pauliny-Toth I.I.K, Williams P.J.S., 1969, ApJ 157, 1
\bibitem[Kulkarni et al 1990]{Ku:90}
Kulkarni V.K., Mantovani F., Pauliny-Toth I.I.K., 1990, A\&AS 82, 41
\bibitem[Reich 1993]{Re:93}
Reich W., 1993, in: Effelsberg News 2, MPIfR Bonn 
\bibitem[Rengelink et al. 1997]{Re:97}
Rengelink R., Tang Y., de Bruyn A.G., Miley G.K., Bremer M.N., R\"ottgering
H.J.A., Bremer M.A.R., 1997, A\&AS 124, 259
\bibitem[Roger et al. 1973]{Ro:73}
Roger R.S., Bridle A.H., Costain C.H., 1973, AJ 78, 1030
\bibitem[Vigotti et al. 1989]{Vi:89}
Vigotti M., Grueff G., Perley R., Clark B.G., Bridle A.J., 1989, AJ 98, 419 
98, 419 
\end{thebibliography}
\end{document}